\documentclass[]{mn2e}
\usepackage{epsfig}

\title[The Progenitor of SN 2002ic]{A single-degenerate model for the
progenitor of the Type Ia supernova 2002ic}

\author[Han \mbox{\rm\&} Podsiadlowski]
{Z.~Han$^1$\thanks{E-mail: zhanwen@public.km.yn.cn}, Ph.~Podsiadlowski$^2$\\
\it 
$^1$ National Astronomical Observatories / Yunnan Observatory, 
the Chinese Academy of Sciences, P.O.Box 110, Kunming, 650011, China\\
$^2$ University of Oxford, Department of Astrophysics, Keble Road, Oxford,
OX1 3RH
}
\newcommand{\Ms}{\mbox{\,M$_{\odot}$}}
\begin{document}
\maketitle

\begin{abstract}
Supernova 2002ic was an atypical Type Ia supernova (SN Ia) with
evidence for substantial amounts of hydrogen associated with the
system. Contrary to previous claims, we show that its unusual
properties can be understood within the framework of one of the most
favoured progenitor models, the so-called supersoft channel. This
requires that the donor star was initially relatively massive ($\sim
3\Ms$) and that the system experienced a delayed dynamical
instability, leading to a large amount of mass loss from the system in
the last few $10^4\,$yr before the explosion. This can produce the
inferred hydrogen-rich circumstellar environment, most likely with a
disc-like geometry. However, to apply these models
requires a larger accretion efficiency onto the white dwarf than is
assumed in present parameterizations.  If this is confirmed, it would
most likely increase estimates for the frequency of the
single-degenerate channel. Based on population synthesis
simulations we estimate that not more than 1 in 100 SNe Ia should
belong to this subgroup of SNe Ia.
\end{abstract}


\section{Introduction}
SN 2002ic was the first Type Ia supernova (SN Ia) for which
circumstellar hydrogen has been detected unambiguously \cite{ham03}.
The detection of hydrogen in a SN Ia has long been considered
one of the cornerstone observations required to help distinguish
between different progenitor models, in particular between
single-degenerate models with a hydrogen-rich donor star \cite{whe73,nom82}
and the double-degenerate merger model where
two CO white dwarfs merge \cite{ibe84,web84}.
However, the amount of circumstellar hydrogen inferred in the case of
SN 2002ic is much larger than one would naively have expected if the
companion star were a slightly evolved star of a few solar masses as,
e.g., in the supersoft scenario (e.g., van den Heuvel et al.\ 1992;
Rappaport, Di\,Stefano \& Smith 1994): estimates range from a minimum
of $\sim 0.5\Ms$ up to $6\Ms$ \cite{wan04,chu04a,uen04,kot04}.
Moreover, from the observed interaction of
the supernova ejecta with the circumstellar medium (CSM) and the
lightcurve one can deduce that this matter must be located within
$10^{17}$\,--\,$10^{18}\,$cm of the supernova and hence must have been
ejected from the pre-supernova system within the last $\sim 10^4\,$yr
(this, however, is strongly dependent on the assumed ejection
velocity).

Spectropolarimetry observations \cite{wan04} suggest that the
supernova exploded inside a dense, clumpy circumstellar environment,
quite possibly with a disk-like geometry.  Subaru spectroscopic
observations \cite{den04} also provide evidence for an interaction of
the SN ejecta with a hydrogen-rich asymmetric circumstellar 
medium\footnote{However, an ejecta-CSM interaction model in which
the CSM is approximately spherically symmetric
was also proposed to explain the Ca II emission features \cite{chu04b}.}.
Interestingly, a few weeks after the explosion, SN 2002ic was almost
identical to several previously observed Type IIn supernovae 
both spectroscopically \cite{den04} and photometrically \cite{woo04}.
This suggests that, while SN 2002ic was clearly a
rare event, it was not a unique one and that there is at least a
sub-class of hydrogen-rich SNe Ia. One of the key questions that has
not yet been answered is whether this requires a progenitor channel
that is completely separate from the bulk of SNe Ia or whether these
are just rare events with properties on the tail of the distribution
of normal SNe Ia.

Hamuy et al.\ \shortcite{ham03} suggested that SN 2002ic may have been 
related to a ``Type 1 1/2'' supernova, a term coined 
by Iben \& Renzini \shortcite{ibe83} to
describe a thermonuclear explosion inside an asymptotic giant branch
(AGB) star when the CO core mass approaches the Chandrasekhar mass. SN
2002ic cannot have been a Type 1 1/2 supernova since almost all of the
hydrogen must have been ejected from the system before the supernova
(i.e., the thermonuclear explosion did not occur inside a
hydrogen-rich AGB envelope). However, from the onset of the interaction of
the supernova with the CSM \cite{woo04} one can deduce
that some hydrogen must have been quite close to the system, within
$\sim 10^{15}\,$cm \cite{woo04,uen04,chu04a}.  
This requires some truly remarkable
fine-tuning since it implies that the last hydrogen was ejected just
before the explosion, indeed within a few decades (!) before the
explosion. Another problem with this scenario is that theoretical
arguments suggest that the hydrogen-rich envelopes in AGB stars are
lost in a superwind long before the CO core has come close to the
Chandrasekhar mass \cite{han94}, except
possibly when the metallicity of the star is extremely low\footnote{SN
2002ic occurred in a dwarf elliptical galaxy and therefore the
metallicity could be quite low. This suggest that determining the
metallicity in the CSM or the stellar neighborhood of SN 2002ic or
some of the other related Type IIn supernovae could provide a useful
test of a SN 1 1/2 scenario.}.

Livio \& Riess \shortcite{liv03} argued that despite the detection of hydrogen,
SN 2002ic could provide evidence in support of a double-degenerate
scenario, suggesting that the supernova was caused by the merger of
two CO white dwarfs inside a hydrogen-rich common envelope (similar to
an earlier suggestion by Sparks \& Stecher 1974), but where the common
envelope was ejected just a few decades before the explosion. In this
model, one would expect a fairly thin shell of hydrogen ejecta,
similar to what is seen in planetary nebulae with close binary
cores. This appears not to be compatible with the large extent of the
hydrogen-rich CSM which ranges from a few $10^{15}\,$cm to at least a
few $10^{17}\,$cm \cite{wan04}. Moreover, Chugai \& Yungelson
\shortcite{chu04a} have pointed out that in order to have a merger within 
100\,yr after the ejection of the common envelope requires a very
tight post-common-envelope orbit with an orbital separation of $\sim
2\times 10^9\,$cm. This, in turn, they argue, implies that the energy
imparted to the envelope ejecta, which is of order the orbital energy
released in the spiral in, would far exceed the kinetic energy deduced
for the CSM from Hamuy et al.\ \shortcite{ham03}.

A perhaps more attractive scenario for SN 2002ic links it to the
symbiotic channel for SNe Ia \cite{hac99a}.  In a
variant of this scenario, the progenitor would have been a fairly wide
binary and the mass donor a massive AGB star or possibly even a Mira
variable, where the CSM originates from matter that was stripped off
the companion star by an energetic wind from the accreting white
dwarf \cite{ham03,den04,chu04a}.  
This requires very efficient stripping to produce
the observationally deduced CSM and again a certain amount of
fine-tuning, but encounters perhaps with the fewest obvious objections.

In this paper, we ask a more conservative question, namely whether the
observed properties of SN 2002ic can be understood within the
framework of the more standard single-degenerate model that links SNe
Ia to supersoft X-ray sources \cite{van92,rap94}, 
where the companion star initially is a
main-sequence star or a slightly evolved star of up to $\sim
3.5\Ms$.  Indeed, as we will show in \S~2, there is a certain
parameter space which may produce systems like SN~2002ic provided that
the accretion efficiency onto the white dwarf is somewhat
increased. The scenario is speculative, though perhaps arguably less
speculative than some of the other proposals for SN~2002ic.
In \S~3 we discuss the implications of this scenario, in
particular what it implies for the overall supernova rates of the
supersoft channel, and suggest some observational tests.

\section{Binary evolution calculations}

Recently, Han \& Podsiadlowski \shortcite{han04} 
performed a systematic study of
the single-degenerate supersoft channel for SN Ia progenitors in which
they carried out detailed binary evolution calculations with the
Eggleton stellar evolution code \cite{egg71,egg72,egg73,han94,pol95} for about
2300 close WD binaries to determine the initial parameters of WD
binaries in the orbital period -- secondary mass ($P_{\rm
orb}$\,--\,$M_2$) plane that can lead to a SN Ia. Since this study was
not able to follow the accretion history of the white dwarf, Han \&
Podsiadlowski (2004) adopted the formalism of 
Hachisu et al.\ \shortcite{hac99a,hac99b}
to determine what fraction of the transferred mass was accreted by the
white dwarf.

A typical supersoft model cannot lead to a supernova like SN~2002ic
since at the time of the explosion the typical mass transfer rate and
consequently systemic mass loss rate tend to be less than
$10^{-6}\Ms\,$yr$^{-1}$, which is far too low to explain the observed
CSM around SN~2002ic (see, e.g., Fig.~1 of Han \& Podsiadlowski 2004
and also Langer et al.\ 2000). However, the detailed mass-transfer
history depends strongly on the initial binary parameters (the initial WD
mass, $M_{\rm WD}^0$, the initial secondary mass, $M_2^0$, and the
initial orbital period, $P_{\rm orb}^0$). To explain systems like
SN~2002ic requires that the mass-transfer rate increases dramatically
to values well in excess of $10^{-4}\,\Ms\,$yr$^{-1}$ when the white
dwarf approaches the Chandrasekhar mass. At such rates, very little
mass can be accreted by the white dwarf and most of it must be lost
from the system. This implies that the WD must have accreted most of
its mass before this phase. This type of evolution is indeed realized
for binary systems near the upper edge of the allowed parameter space
in the $P_{\rm orb}$\,--\,$M_2$ plane (see, e.g., Figs.~1 and ~2 of Han \&
Podsiadlowski 2004), where systems are close to
experiencing a delayed dynamical instability.

\subsection*{Delayed dynamical instability}

For a given initial orbital period and initial white dwarf mass, the
maximum initial secondary mass is determined by the condition that
mass transfer must be dynamically stable. This depends mainly on the
mass ratio.  For radiative stars, the critical mass ratio is typically
$\sim 3$ (see, e.g., Podsiadlowski, Rappaport \& Pfahl 2002 [PRP]; Han
\& Podsiadlowski 2004). For radiative stars, however, it is well known
that this instability is significantly delayed (Hjellming \& Webbink
1987; PRP) and that substantial mass transfer can occur before its
onset.

\begin{figure}
\centerline{\epsfig{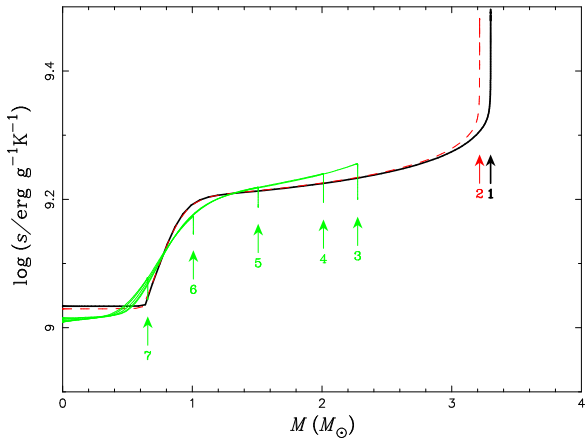}}
\caption{Evolution of the specific entropy profile of the mass donor 
as a function of time for the binary calculation shown in Figures \ref{hrd} 
and \ref{mdot}.
The thick solid curve shows the profile at the onset of mass transfer. 
The mass transfer rate starts to run away when the high entropy
outer region has been lost, and the subsequent evolution of the star
is adiabatic (thin solid curves).}
\label{entropy}
\end{figure}

This behaviour can be understood from the evolution of the entropy
profile of the donor star. A radiative star has a strongly rising
specific entropy in the outer layers of the star (see Fig.~\ref{entropy}). 
As long
as the star has this entropy spike, it generally can adjust its radius
to fit whatever the changing Roche-lobe radius requires. But when the
flatter entropy profile in the inner region is exposed, the star starts
to behave more like a convective star (which has a flat entropy
profile) which tend to expand when mass is taken off adiabatically. At
this point, the mass transfer rates start to increase dramatically and
may even run away, i.e., lead to dynamical mass transfer. The delay
time depends on the mass of the star but is typically of order
$10^5\,$yr for a star of several \Ms\ (see, e.g., Fig.~12 of PRP), in
which a significant amount of mass can be transferred.

\begin{figure}
\centerline{\epsfig{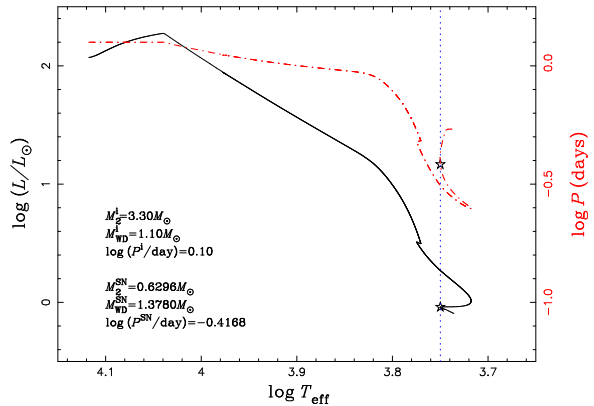}}
\caption{Luminosity of the mass donor (solid curve, left axis) and
binary orbital period (dot-dashed curve, right axis) as a function of
effective temperature for a binary evolution calculation applicable to
SN 2002ic.  The initial masses of the mass donor and white dwarf are
3.3\Ms\ and 1.1\Ms, respectively, and the initial orbital period is
$30\,$hr. At the time of the explosion
(indicated by a vertical dotted line), the secondary has a mass of
0.63\Ms\ and the orbital period is $\sim 9\,$hr.}
\label{hrd}
\end{figure}

\begin{figure}
\centerline{\epsfig{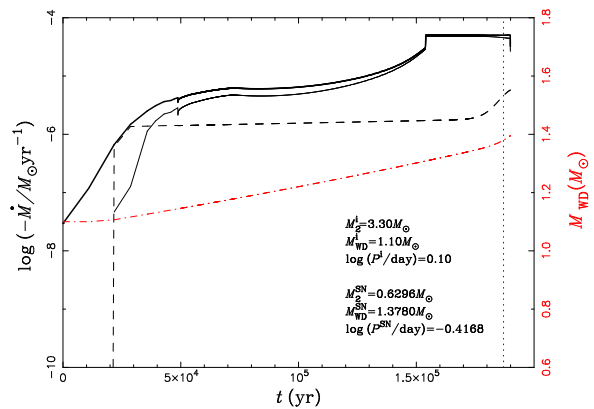}}
\caption{The evolution of the mass transfer rate (thick solid curve),
mass loss rate from the system (thin solid curve) and WD accretion
rate (dashed curve) as a function of time for the binary calculation
in Figure~\ref{hrd} (left axis).  Note the high mass-loss rate in the last
$3\times 10^4\,$yr before the explosion (limited to $0.5\times
10^{-4}\Ms\,$yr$^{-1}$). The dot-dashed curve shows the evolution of
WD mass (right axis). The vertical dotted curve again indicates
the time of the explosion.}
\label{mdot}
\end{figure}

Figures~\ref{hrd} and \ref{mdot} show a binary evolution calculation 
which may be applicable to
SN~2002ic. The WD binary has an initial orbital period of $30\,$hr and
initially consists of a 1.1\Ms\ CO WD and a somewhat evolved secondary
of 3.3\Ms, which has a central hydrogen mass fraction of
0.3 at the beginning of mass transfer.  
Mass transfer initially occurs on a thermal timescale, but is
dynamically stable; the mass transfer rate (thick solid curve) rises
quickly over the first $5\times 10^4\,$yr to $\sim 5\times
10^{-6}M_\odot{\rm yr}^{-1}$, 
well in excess of the rate the white dwarf can accrete
(dashed curve of Fig.~\ref{mdot}), and most of the
transferred mass is lost from the system 
(thin solid curve of Fig.~\ref{mdot}).  The mass
transfer continues to grow more slowly for the next $10^5\,$yr. At
this point, the secondary has completely lost the high entropy spike
(see the profile marked `3' in Fig.~\ref{entropy}), and the
system starts to encounter the delayed dynamical instability, where
the radius of the secondary can no longer adjust to just fill its
Roche lobe radius, and the mass transfer rate increases dramatically.
We cannot follow the subsequent evolution of the system properly and
therefore fixed the maximum transfer rate at $0.5\times
10^{-4}\Ms\,$yr$^{-1}$, assuming that the system would survive as a
mass-transferring binary (see \S~3.2 for further discussion). This
means that in our calculations the secondary was allowed to overfill
its Roche lobe by a moderate amount (the formal radius of the
secondary at the time of the explosion exceeded the Roche lobe by
$\sim 17\%$). When the system encounters the delayed
dynamical instability, the WD mass has grown to 1.31\Ms. 
It continues to grow until it explodes as a SN Ia $\sim
3\times 10^4\,$yr later. In this final phase, the system is expected
to lose mass at a rate well in excess of $10^{-4}M_\odot {\rm yr}^{-1}$ 
(the total mass
lost in this phase is $\sim 1.5\,$\Ms).  At the time of the explosion
the orbital period is only 9\,hr and the secondary has a mass of
0.63\Ms.

One of the key points in this scenario is that the time to the
encounter of the delayed dynamical instability is completely
determined by the initial entropy profile of the secondary and the
evolution of the binary parameters, which in turn depends strongly on
the angular momentum loss associated with the mass loss from the
system. This time cannot significantly exceed $2\times 10^5\,$yr
 and this is the only time
the WD has to grow to reach the Chandrasekhar mass. Whether this
scenario can produce a SN Ia, therefore depends both on the initial
white dwarf mass (more massive white dwarfs are favoured, since it is
then easier to reach the Chandrasekhar mass) and most importantly on
the accretion efficiency. In order to be able to grow the WD mass
sufficiently in the above calculation, we had to {\em arbitrarily
increase} the accretion efficiency by a factor of 2.5 over the value
assumed in our standard model (i.e. the model based on the efficiency
calculations by Hachisu et al.\ 1999b). Whether this is at all
reasonable or not will be discussed further in the next section.

\section{Discussion}

\subsection{The accretion efficiency}

\begin{figure*}
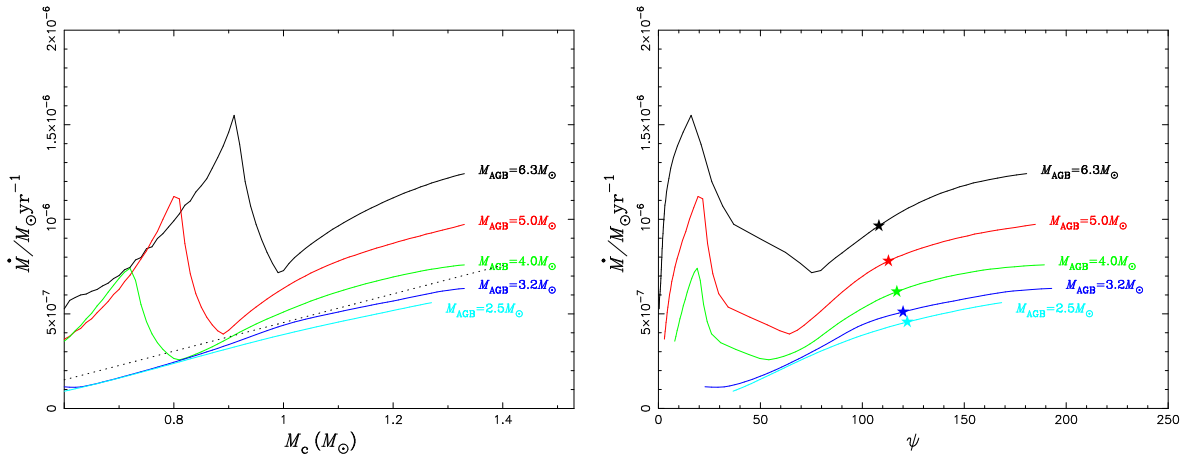

\centerline{\epsfig{file=z02a.ps,angle=270,width=7.7cm}\ \
\epsfig{file=z02b.ps,angle=270,width=7.7cm}}
\caption{Core growth rate for different ($Z=0.02$) AGB stars with
masses from 2.5 to 6.3\Ms (as indicated) without convective overshooting
as a function of core mass ({\em left panel}) and as a function of the
central degeneracy parameter $\psi$ ({\em right panel}).  The dotted
curve shows the critical (i.e. maximum) accretion rate of
hydrogen-rich material onto a CO WD from Hachisu et al.\ (1999b).  
The stars in the right panel
indicate when the core mass is $1.1M_\odot$.}
\label{z02}
\end{figure*}

One of the major physical uncertainties in the scenario for SN~2002ic
presented here is whether the efficiency for accretion onto the white
dwarf is indeed as large as required (2.5 times the value assumed in
our canonical model). The accretion efficiency is in general quite
uncertain and depends on rather uncertain physics, e.g., the
occurrence of nova explosions and helium flashes and the associated
mass loss, the details of a disc wind that is assumed to carry away
the excess mass that cannot be accreted \cite{kat94}, etc.
In particular, none of the present estimates take into account the
role of the rotation of the accreted material. Yoon \& Langer \shortcite{yoo04}
recently showed for the case of helium-accreting white dwarfs that
rotation dramatically changes the evolution of the white dwarf: it
generally reduces the violence of helium flashes, it may even suppress
them, and hence is expected to increase the accretion efficiency and
the parameter range for which white dwarfs can grow efficiently.  On
the other hand, the critical explosion mass may increase well above
the standard Chandrasekhar mass (up to $\sim 1.8\Ms$; Yoon \& Langer
2005), i.e., require even more mass to be accreted. Whether the net
effect is to increase the efficiency to the value required here is
not clear at the moment.

However, we can at least make a plausibility check by comparing the
``accretion rates'' (more precisely the core growth rates) of the CO
cores in AGB stars with the rate required for SN~2002ic. 
These are shown in Figure~\ref{z02} for AGB stars from 2.5 to 6.3\Ms, 
both as a function of core mass and central degeneracy parameter.
Note that real AGB stars do not reach such high
CO core masses as shown in the figure since they eject their envelopes 
before. These calculations
were done without mass loss in order to be able to estimate the core 
accretion rate. They do, however, remain CO cores for the masses shown and 
would ultimately ignite carbon in the center leading to a thermonuclear
runaway (sometimes referred to as a 1 1/2 supernova). Only the more
massive AGB stars ignite carbon off-center and are converted into
ONeMg stars (as, e.g., discussed in Iben 1974).
Our integration on the AGB uses relatively large timesteps which
suppresses thermal pulsations; therefore the accretion rate on the CO
core is only correct in a time-averaged sense. Moreoever, this
exercise is certainly not a proof of higher accretion efficiencies;
this still requires much more detailed work, as discussed earlier.
As seen from Figure~\ref{z02}, the core growth rates
varies by almost a factor of 3 in the region of interest (i.e., above
a core mass of 1\Ms) and can be larger by more than of a factor of 2
than the critical rate, shown as a dotted curve in the left panel,
used in the Hachisu et al.\ \shortcite{hac99b} parameterization.  The highest
rates are comparable to the rates required in the scenario
presented above, suggesting that the assumption of a higher accretion
efficiency is at least not {\em a priori} unreasonable.

\subsection{The pre-supernova mass loss}

A second uncertainty in this scenario is related to the final
mass-loss phase which we cannot treat properly with our code, since
multi-dimensional and possibly hydrodynamical effects start to become
important.  In our calculation, the secondary overfilled its Roche
lobe by about 49\% at its most extreme; this is
likely to lead to the formation of a common-envelope (CE) phase
surrounding both binary components.  Whether this leads to a dramatic
spiral-in and the ultimate merger of the components or not is not so
clear. At least initially the envelope will be in co-rotation with the
binary, which implies that at least initially there is no friction to
extract energy from the orbit and drive a spiral-in. Moreover, if the
density in the CE is low (as is, e.g., expected to be the case for
donors with radiative envelopes; Podsiadlowski 2001), only a moderate
shrinking of the orbit is required to inject enough energy into the CE
to eject the excess mass. In this case, a moderate orbital shrinking
may drive a {\em frictionally driven wind}\footnote{Such frictionally
driven orbital evolution was not included in our binary calculation.}.
Because of the large rotation of the system, this wind would be
expected to be very anisotropic, most likely resemble an equatorial
outflow (perhaps confined to within 30$^{\rm o}$ of the orbital plane
as in the case of SS 433; Blundell et al.\ 2001) and have a velocity
of order or somewhat lower than the orbital velocity of the
binary. This is consistent with the inferred measurement of the wind
velocity from the H$\alpha$ P-Cygni profile of $\sim 100$\,km\,s$^{-1}$
\cite{kot04}. We also note that there may be mass loss
through one or both of the outer Lagrangian points (L2 and/or
L3). This could produce a circumbinary, disc-like structure that may
affect the spectral evolution of the supernova ejecta (see Mazzali et
al.\ 2005). It is also possible that there exists a bipolar outflow,
though it is not clear whether this would affect the SN interaction.

There is, however, another interesting possibility, namely that the
spiral-in runs away and leads to the complete merging of the CO WD
with the secondary. Once the CO WD has settled at the centre of the
merger product, the envelope will expand to red-giant dimensions and
the object resemble an AGB star, except that it already has a rather
massive CO core of $1.35\Ms$; the final product will be {\em an object
that may never be able to evolve from a single star} (see the
discussion in \S~1). In this case, it may be easier for the core to
grow to reach the Chandrasekhar mass. Moreover, the final phase is
expected to be accompanied by a superwind phase 
(note, it is not a classical superwind in an AGB star that is
driven by MIRA pulsations, but the mass loss must be so high to
resemble a superwind), possibly producing
the type of CSM inferred for SN~2002ic.  In this case, SN~2002ic could
indeed be related to a SN 1 1/2 \cite{ham03}, but avoid many
of the problems of a single-star scenario.

\subsection{Frequency estimates}

\begin{figure*}
\centerline{\epsfig{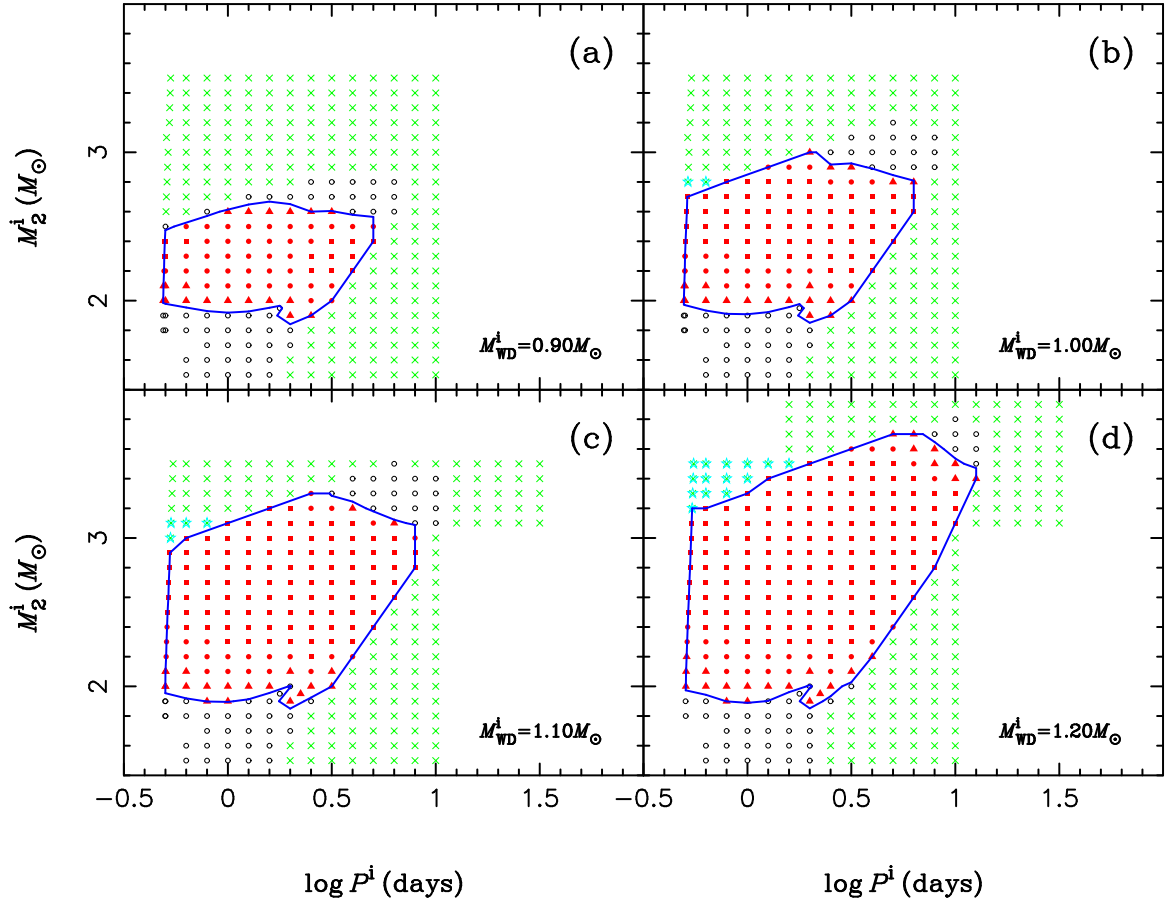}}
\caption{Estimated parameter regions for a delayed dynamical instability
in the orbital period -- secondary mass ($\log P_{\rm orb}\,$--$\,M_2$) 
plane for different initial WD masses, $M_{\rm WD}$, as
indicated. The solid contours enclose the parameter regions for
which a WD binary is expected to explode as a SN Ia in the canonical
model. The stars indicate the parameter regions for a delayed
dynamical instability, as may be applicable to SN~2002ic.}
\label{grid}
\end{figure*}

Supernovae similar to SN 2002ic clearly form a relatively rare 
subclass of SNe Ia; no more than perhaps 1 in 100 SNe Ia can be of this
type \cite{ham03,den04,chu04b,woo04}.  
Therefore the evolutionary channel leading to a
SN Ia has to be a rare one, and a certain amount of fine-tuning is not
only acceptable, but is required. In the delayed dynamical
instability scenario, the white dwarf has to be quite massive
initially ($\ga 1.0\Ms$), since the amount of mass that can be accreted
before the dynamical instability sets in is limited. This also implies
that the secondary mass has to be relatively large initially ($\ga
3\Ms$), since the critical mass ratio for a delayed dynamical
instability is $\sim 3$.  Moreover, at the time of the instability the
WD mass has to be already close to the Chandrasekhar mass ($\sim
1.35\Ms$), since very little mass can be accreted subsequently. To
estimate the frequency for our scenario, we did not perform another
large series of binary calculations with an increased accretion
efficiency. Instead, we re-evaluated the outcome for our 2300
sequences from Han \& Podsiadlowski \shortcite{han04}, examining which of our
sequences would fulfil the above constraints for SN~2002ic if the
accretion efficiency were increased by a factor of 2.5. The results of
this exercise are shown in Figure~\ref{grid}, which shows the expected fate in
the orbital period -- secondary mass ($\log P_{\rm orb}\,$--$\,M_2$)
plane for different initial WD masses. The stars indicate sequences
that have the correct properties for SN~2002ic. Altogether 18 out of
our 2300 sequences satisfy these constraints. 

\begin{figure*}
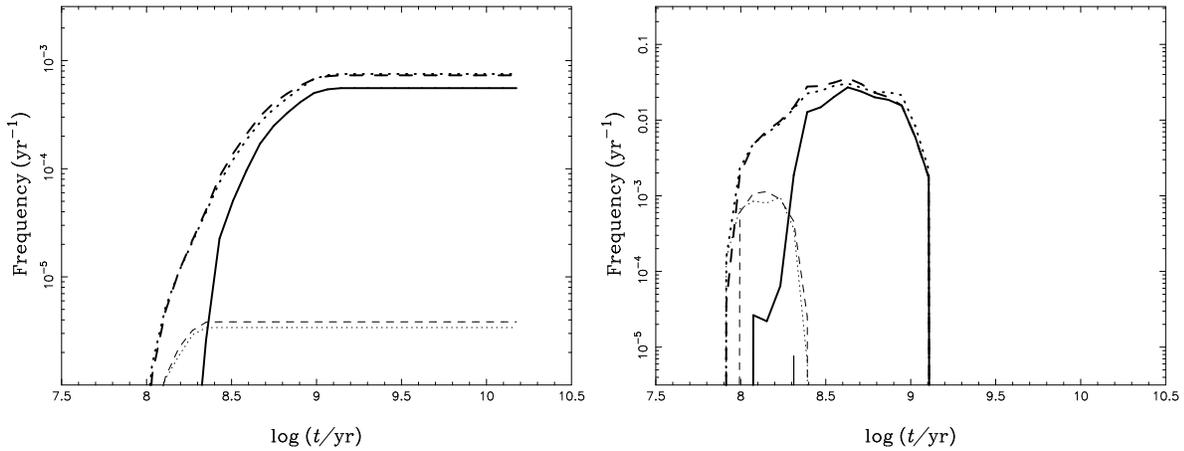

\centerline{\epsfig{file=const.ps,angle=270,width=7.7cm}\ \
\epsfig{file=single.ps,angle=270,width=7.7cm}}
\caption{{\em Left panel:} the evolution of birthrate of SNe Ia for a
constant Pop I star-formation rate (3.5\Ms\,yr$^{-1}$). The thick
curves are for a ``normal'' SNe Ia, while the thin ones curves are for
SNe Ia from the ``delayed dynamical'' channel. The solid, dashed and
dotted curves are for different CE-ejection parameters, $\alpha_{\rm
CE} = \alpha_{\rm th} = 1.0, 0.75, 0.5$, respectively (see Han \&
Podsiadlowski 2004 for further details). {\em Right panel:} similar to
the left panel, but for a single star burst of $10^{11}\Ms$.}
\label{rate}
\end{figure*}
  
We used these constraints to perform full binary population synthesis
calculations to estimate the frequencies of supernovae of this
type. Figure~\ref{rate} compares the evolution of birthrates for a constant
star formation rate (of 3.5\Ms\,yr$^{-1}$; left panel) and a single
star burst (right panel) both for a typical SN~Ia (thick curves) and a
delayed dynamical scenario (thin curves), where we varied the binary
population synthesis parameters over a reasonable range (see Han \&
Podsiadlowski 2004 for further discussion). The main result of these
simulations is that the expected overall rate for a delayed dynamical
scenario is about a factor of 200 lower than the overall SN Ia rate
(see the left panel of Fig.~\ref{rate}), although at early times (within $\sim
2\times 10^8\,$yr) of a star burst it is lower by only a factor of 10
or less. These estimates are consistent with the rarity of observed
supernovae similar to SN 2002ic and independent estimates. This type of
SNe Ia could therefore be used to trace star formation, with a time delay 
somewhat longer than for core collapse supernovae and shorter than
for the bulk of ``normal'' SNe Ia, of which these form the youngest
sub-class.

\section{Conclusions}

The most important conclusion of this study is that SN~2002ic, even
though it clearly had an atypical progenitor evolution, can still be
understood within the framework of the arguably most favoured
single-degenerate channel which links SNe Ia to supersoft X-ray
sources. Even within this channel, progenitors display substantial
diversity, and this may be able to account at least in part for the
observed diversity of SNe Ia. SN~2002ic may represent an extreme case
with a donor star near the upper mass ($\sim 3 - 3.5\Ms$) allowed in
this channel and may have experienced a delayed dynamical instability
where the mass loss from the system increased dramatically just in the
last few $10^4\,$yr before the explosion. If this is the case,
SN~2002ic may in fact provide useful constraints on the supersoft
channel and help to calibrate some of the important physical input
parameters. In particular, as we have shown here, this model requires
a significantly larger accretion efficiency (by at least a factor of
2) than assumed in the parameterization by Hachisu et al.\ \shortcite{hac99b}.
Another implication of this is that it most likely suggests that the
parameter range for the initial binary parameters that can lead to a
SN Ia is much larger than in previous studies (e.g.  Han \&
Podsiadlowski 2004; Fedorova, Tutukov \& Yungelson 2004), increasing
estimates for the frequency of this channel and making them more
consistent with the observed frequency, thereby alleviating one of the
major objections to it (see e.g. Fedorova et al.\ 2004).

SNe Ia like SN~2002ic should be rare events, since the parameters of
the progenitor systems are very restricted. We estimate that not more
than 1 in 100 SNe Ia should fall into this subclass. Since they
require an intermediate-mass secondary, they should only be found in
stellar populations with relatively recent star formation (with the
last $\sim 3\times 10^8\,$yr). Unlike a SN 1 1/2, the system still has
a companion star at the time of the explosion which may interact with
the supernova ejecta. Another potential test of the model is the
amount of hydrogen found in the surrounding circumstellar medium
(CSM). In this model, the maximum mass is limited to the mass of the
companion star and the mass in the nearby CSM cannot reasonably exceed
$\sim 2\Ms$, which is substantially less than some of the more
extreme estimates at present (see \S~1). In addition, one may expect
that most of this material should form a disk-like outflow
(or possibly even a circumbinary disc).

\section*{Acknowledgements}
We are grateful to an anonymous referee for his/her useful comments.
This work was in part supported by a Royal Society UK-China Joint Project
Grant (Ph.P and Z.H.), the Chinese Academy of Sciences under
Grant No.\ KJCX2-SW-T06, and Natural Science Foundation of
China under Grant Nos.\ 10433030 and 10521001 (Z.H.) and
a European Research \& Training Network on Type Ia Supernovae
(HPRN-CT-20002-00303).

\end{document}